\documentclass[conference]{IEEEtran}
\IEEEoverridecommandlockouts

\usepackage{cite}
\usepackage{amsmath,amssymb,amsfonts}
\usepackage{algorithmic}
\usepackage{graphicx}
\usepackage{textcomp}
\usepackage{xcolor}

\usepackage[T1]{fontenc}
\usepackage[utf8]{inputenc}

\usepackage[compatibility=false]{caption}
\usepackage{subcaption}
\usepackage[percent]{overpic}

\def\BibTeX{{\rm B\kern-.05em{\sc i\kern-.025em b}\kern-.08em
    T\kern-.1667em\lower.7ex\hbox{E}\kern-.125emX}}

\begin{document}

\title{Structured Clifford+T Circuits for Efficient Generation of Quantum Chaos} 
\author{\IEEEauthorblockN{Asim Sharma}
\IEEEauthorblockA{\textit{Computer Science Department} \\
\textit{Missouri University of Science and Technology}\\
Rolla, Missouri, USA \\
asn32@mst.edu}
\and
\IEEEauthorblockN{Avah Banerjee}
\IEEEauthorblockA{\textit{Computer Science Department} \\
\textit{Missouri University of Science and Technology}\\
Rolla, Missouri, USA \\
banerjeeav@mst.edu}
}

\maketitle

\begin{abstract}
We investigate the emergence of quantum chaos and unitary T--design behavior in derandomized Clifford+$T$ circuits using causal cover architectures. Motivated by the need for deterministic constructions that can exhibit chaotic behavior across diverse quantum hardware platforms, we explore deterministic Clifford circuit architectures (random Clifford circuits with causal cover, bitonic sorting networks, and permutation-based routing circuits) to drive quantum circuits toward Wigner--Dyson (WD) entanglement spectrum statistics and OTOC decay. Our experiments demonstrate that causal connectivity, not circuit depth or randomness, is a critical feature that drives circuits to chaos. We show that initializing with $n$ T--states and adding a second T--layer after a causally covered Clifford evolution yields consistent OTOC decay and WD statistics. This also enables deeper understanding of the circuit structures that generate complex entanglement behavior. Notably, our work suggests polylogarithmic--depth deterministic circuits suffice to approximate chaotic behavior, highlighting that causal connectivity is sufficient for operator spreading to induce Wigner--Dyson entanglement statistics and OTOC decay.
\end{abstract}


\section{Introduction}
Quantum chaos has been widely studied both as a fundamental phenomenon in many--body quantum systems and as a potential resource for quantum computational advantage. Despite the breadth of research, a formal and universally accepted benchmark for chaos in quantum circuits is still lacking. Among the most commonly adopted indicators is the Out--of--Time--Order Correlator (OTOC), which measures the degree to which initially commuting operators fail to commute after evolution, reflecting information scrambling. Another key concept is the unitary design \( T \), which compares the \( t \) th moments of the unitary distribution of a circuit with those of the Haar measure. Since the Haar measure defines the maximally chaotic ensemble of unitaries, achieving a higher \( T \)--design indicates stronger chaotic properties.

Recent work, such as in \cite{zhou2020single}, has shown that applying a single \( T \)--gate after an entanglement heating(which increases entanglement entropy) layer composed of Clifford gates leads the circuit’s entanglement spectrum statistics to move closer to the Wigner--Dyson (WD) distribution. This suggests that each additional \( T \)--gate pushes the circuit closer to forming a unitary \( T \)--design, enhancing its chaotic character. Moreover, the study in \cite{mi2021information} highlights that when the unitary used in an OTOC computation exhibits significant operator entanglement, the resulting OTOC decay mirrors that of chaotic dynamics, providing further evidence of chaos.

Our research has three main motivations. First, we aim to design deterministic quantum circuits that consistently exhibit chaotic behavior and can be implemented across different quantum hardware platforms. Second, we seek to explore the mechanisms behind the emergence of chaos and complex entanglement in structured circuits. Finally, we investigate how nonstabilizer character (or \( T \) --ness) introduced via T--gates propagates through the circuit and contributes to chaotic unitary statistics. By leveraging structured architectures with, we aim to provide a general method for generating chaos in all types of quantum hardware.

\begin{figure*}[htbp]
  \centering
  \begin{subfigure}[t]{0.18\textwidth}
    \centering
    \includegraphics[width=\linewidth]{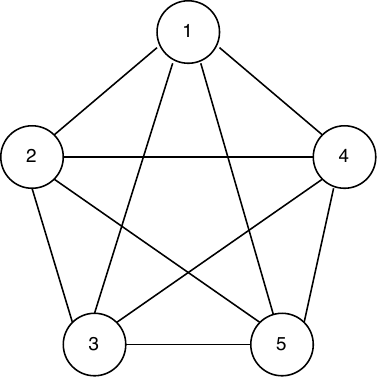}
    \caption{} 
    \label{fig:graph}
  \end{subfigure}
  \hfill
  \begin{subfigure}[t]{0.18\textwidth}
    \centering
    \includegraphics[width=\linewidth]{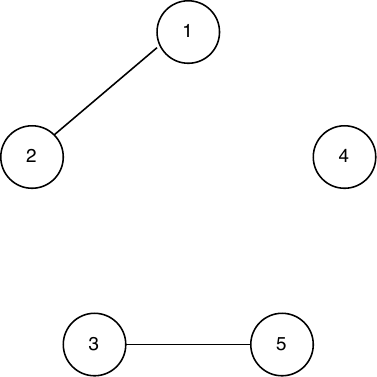}
    \caption{}
    \label{fig:t1}
  \end{subfigure}
  \hfill
  \begin{subfigure}[t]{0.18\textwidth}
    \centering
    \includegraphics[width=\linewidth]{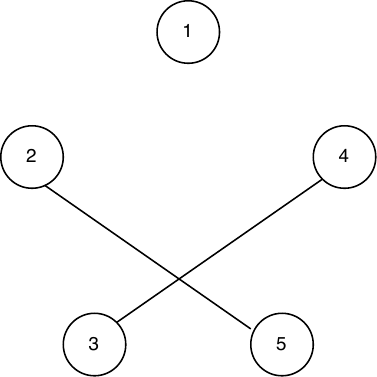}
    \caption{}
    \label{fig:t2}
  \end{subfigure}
  \hfill
  \begin{subfigure}[t]{0.18\textwidth}
    \centering
    \includegraphics[width=\linewidth]{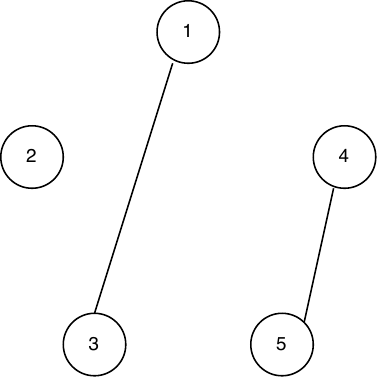}
    \caption{}
    \label{fig:t3}
  \end{subfigure}
  \hfill
  \begin{subfigure}[t]{0.18\textwidth}
    \centering
    \includegraphics[width=\linewidth]{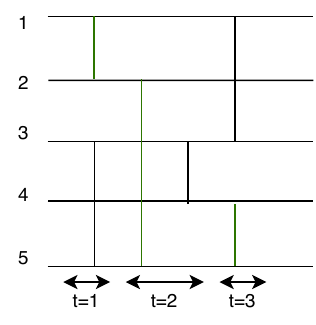}
    \caption{}
    \label{fig:circuit}
  \end{subfigure}
  \hfill

  \caption{
Causal cover: (a) Graph $G = (V, E)$; (b), (c), and (d) represent subgraphs $G_1$, $G_2$, and $G_3$ of $G$, where each $E_i$ (the edge set of $G_i$) is a matching in $G$; (e) shows the resulting circuit constructed from subgraphs $G_1$, $G_2$, and $G_3$. In the circuit, a connection $(u, v)$ is applied at time step $t = i$ if $(u, v) \in E_i$.$(1,4)$ is causally covered here but the circuit as a whole is not causally covered since $(4,2)$ is not causally covered.
}
  \label{fig:causalExplain}
\end{figure*}

\section{Preliminaries}

Quantum chaos in many--body systems is characterized by the emergence of random matrix behavior in otherwise unitary and deterministic dynamics. A central diagnostic of chaos is the out--of--time--order correlator (OTOC) \cite{harrow2021separation,swingle2016measuring}, which captures the sensitivity of a quantum system to perturbations. In our context, the OTOC is defined through the correlator between an operator \( {W}(t) \) and operator \( {V} \) acting on a different qubit:
\begin{equation}
C(t) = \langle W ^\dagger(t) V ^\dagger W(t) {V} \rangle.
\end{equation}
Here, \( \langle \cdot \rangle \) denotes an expectation over the quantum state and $W(t)=\hat{U}(t)^\dagger W \hat{U}(t)$ is the Heisenberg evolution of the operator. This quantity reflects how rapidly information spreads.
To measure \( C(t) \), we use interferometric protocol \cite{swingle2016measuring}. We consider a quantum system \( S \), consisting of qubits \( Q_1 \) through \( Q_n \), along with an ancillary control qubit \( C \). The control qubit is initialized in the state \( |+X\rangle = (|0\rangle + |1\rangle)/\sqrt{2} \). The system then undergoes the following sequence of operations:

\begin{enumerate}
    \item \( I_S \otimes |0\rangle\langle 0|_C + V_S \otimes |1\rangle\langle 1|_C \),
    \item \( U(t)_S \otimes I_C \),
    \item \( W_S \otimes I_C \),
    \item \( U(-t)_S \otimes I_C \),
    \item \( V_S \otimes |0\rangle\langle 0|_C + I_S \otimes |1\rangle\langle 1|_C \)
\end{enumerate}

\noindent to prepare the state:
\[
\frac{(VW_t |\psi\rangle_S)|0\rangle_C + (W_t V |\psi\rangle_S)|1\rangle_C}{\sqrt{2}}.
\]

Finally, the control qubit is measured in the \( X \) bases to extract the real component of C(t). The full correlator is then given by:
\begin{equation}
F = \langle X_C \rangle + i \langle Y_C \rangle.
\end{equation}
We use only real part of F for our experiments. Also, we use different pauli operator as $W$ and $V$ in different circuits, as OTOC should decay for any choice of operators. 

While OTOC measures the scrambling of quantum information, we are also interested in how closely a circuit's unitary distribution approximates the Haar measure. Specifically, a distribution that matches the first \( t \) moments of the Haar measure is known as a \textit{unitary \( t \)--design}. 

Formally, a finite ensemble of unitaries \( X \subset U(d) \), where \( U(d) \) denotes the group of \( d \times d \) unitary matrices, is called a unitary \( t \)-design \cite{roy2009unitary ,zhu2017multiqubit} if:

\[
\frac{1}{|X|} \sum_{U \in X} U^{\otimes t} \otimes (U^*)^{\otimes t} = \int_{U(d)} U^{\otimes t} \otimes (U^*)^{\otimes t} \, d\mu(U)
\]

\noindent where, \( d\mu(U) \) is the Haar measure on \( U(d) \), the unique unitarily invariant probability measure normalized such that \( \int_{U(d)} d\mu(U) = 1 \).

This definition implies that any polynomial of degree at most \( t \) in the matrix elements of \( U \) and \( U^* \) has the same expectation when averaged over \( X \) as when averaged over the Haar measure. Thus, constructing circuits that form or approximate a unitary \( t \)--design allows us to mimic the statistical properties of truly random unitaries, which are central to understanding quantum chaos and thermalization. While the Clifford group can successfully create unitary 3--design, it is proven that clifford group cannot create unitary $t$--design for $t>3$ \cite{zhu2016clifford} \cite{webb2015clifford}. This is why we need non--stabilizer resources(t--gates) to get chaotic distribution.

One commonly used benchmark for identifying \( t \)--design behavior is to compare the entanglement spectrum statistics of the circuit output state with the Wigner--Dyson (WD) distribution from random matrix theory \cite{zhou2020single}. Circuits composed solely of Clifford gates do not yield WD statistics, non--stabilizer resources are must for a circuit to demonstrate WD distribution. In our experiment, we quantify this using the modified level spacing ratio \( \tilde{r} \) as done in  \cite{zhou2020single}, which transitions from Poisson--like behavior to WD--like behavior as the circuit approaches a \( t \)--design. Specifically, \( \tilde{r} \) is computed as
\[
\tilde{r}_k = \frac{\min\{\delta_k, \delta_{k+1}\}}{\max\{\delta_k, \delta_{k+1}\}}, \quad \delta_k = \lambda_{k-1} - \lambda_k,
\]
where \( \lambda_k \) are the ordered eigenvalues of the reduced density matrix obtained by bipartitioning the system. For Poisson statistics, \( \langle \tilde{r} \rangle \approx 0.39 \), whereas for GUE (WD statistics), \( \langle \tilde{r} \rangle \approx 0.6 \). This metric serves as our primary probe for identifying chaotic behavior and \( t \)--design formation.
\begin{figure*}[!t]
  \begin{subfigure}[t]{0.32\textwidth}
    \includegraphics[width=\linewidth]{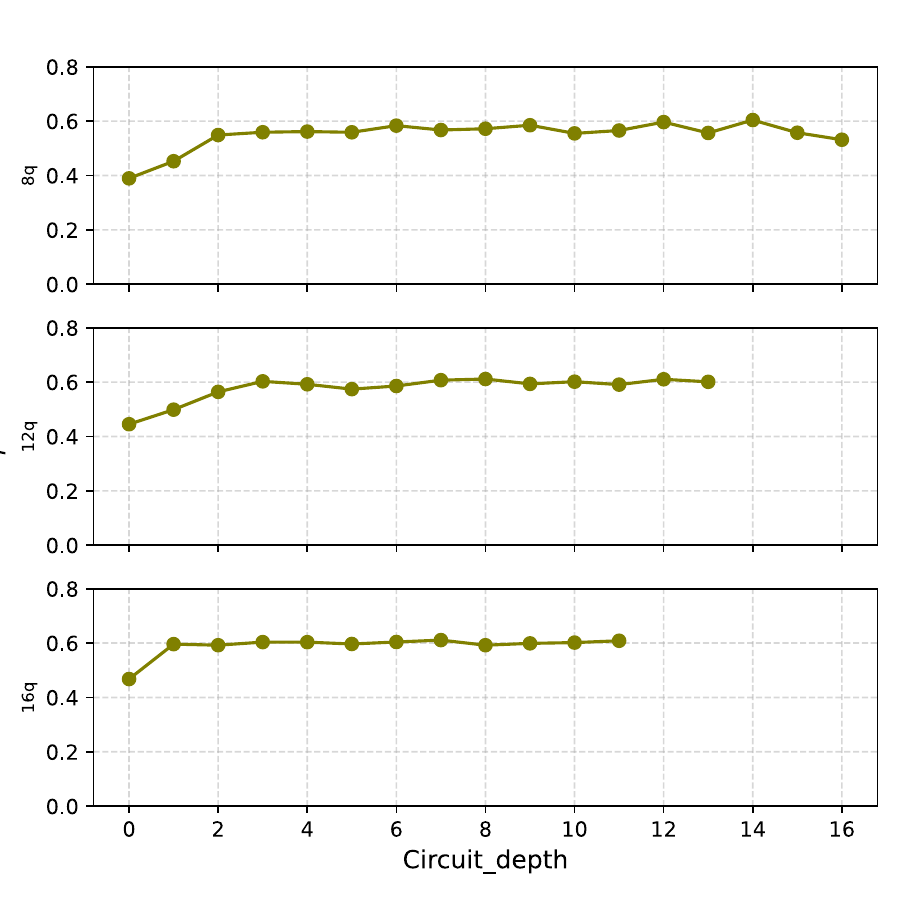}
    \caption{}
    \label{fig:causal_a}
  \end{subfigure}
  \hfill
  \begin{subfigure}[t]{0.32\textwidth}
    \includegraphics[width=\linewidth]{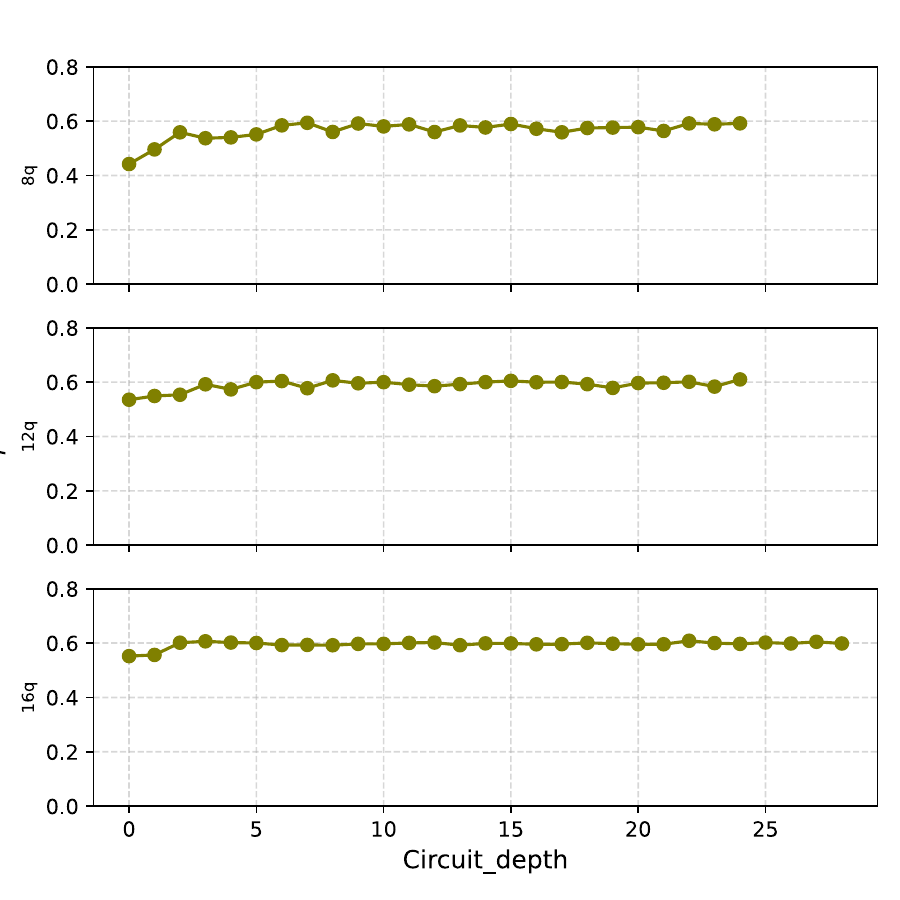}
    \caption{}
    \label{fig:causal_b}
  \end{subfigure}
  \hfill
  \begin{subfigure}[t]{0.32\textwidth}
    \includegraphics[width=\linewidth]{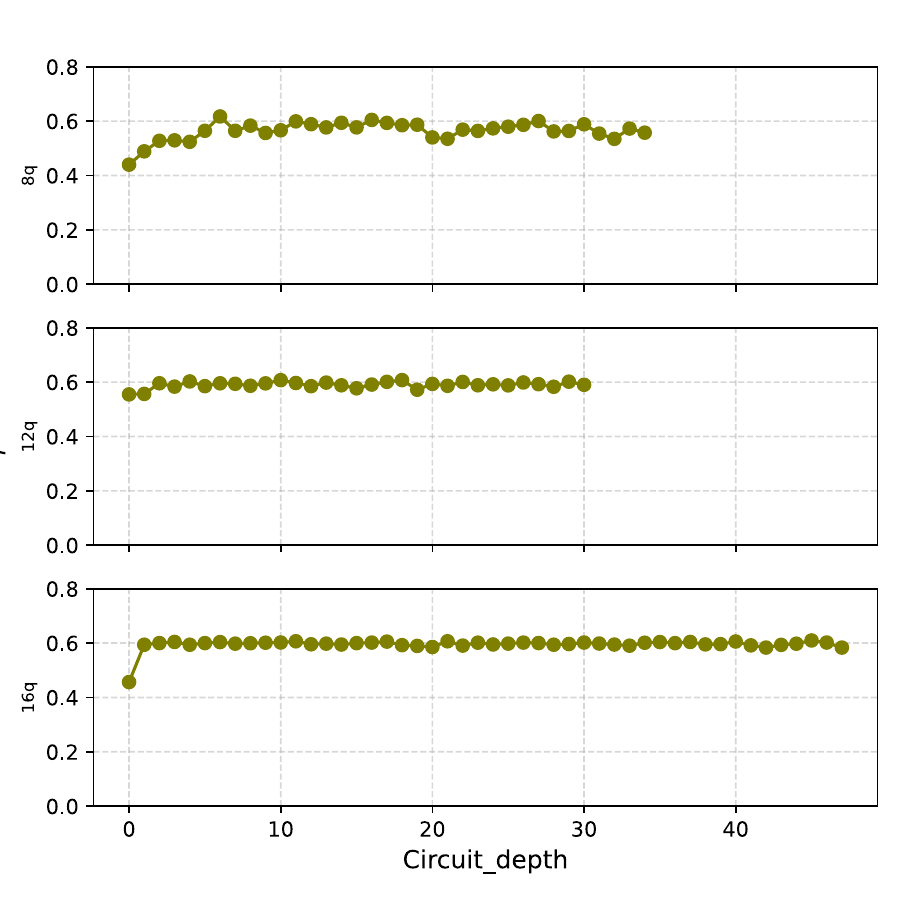}
    \caption{}
    \label{fig:causal_c}
  \end{subfigure}
 \caption{\(\tilde{r}\) measured after the second \(T\)--gate layer. The circuit is initialized with \(n\) \(T\)--states, followed by a Clifford block, a second layer of \(n\) \(T\)--gates, and a final Clifford block. The entanglement heating section (Clifford block) consists of a random Clifford unitary that satisfies the causal cover condition. Subfigures (a), (b), and (c) correspond to increasing depths of this section: (a) \(1\times\) causal cover, (b) \(2\times\) causal cover, and (c) \(3\times\) causal cover. \(\tilde{r}\) is averaged over 15 different circuits for each depth.}
  \label{fig:causal_cover}
\end{figure*}

\begin{figure*}[!t]
  \begin{subfigure}[t]{0.32\textwidth}
    \includegraphics[width=\linewidth]{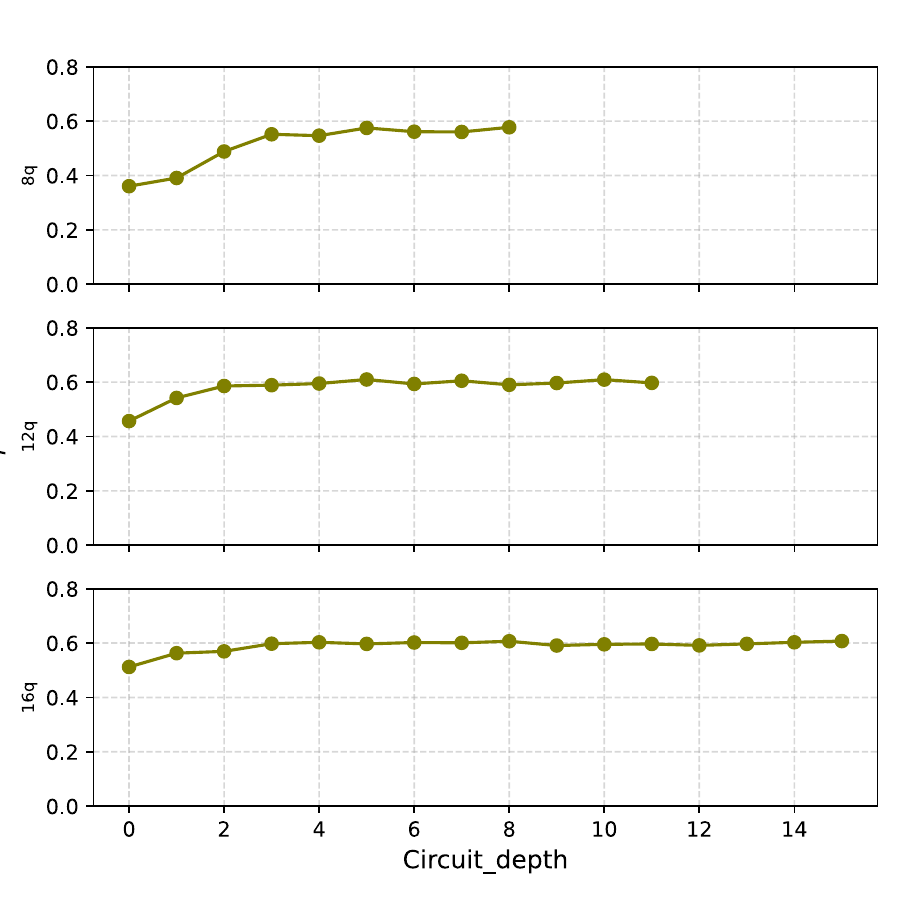}
    \caption{}
    \label{fig:causal_1}
  \end{subfigure}
  \hfill
  \begin{subfigure}[t]{0.32\textwidth}
    \includegraphics[width=\linewidth]{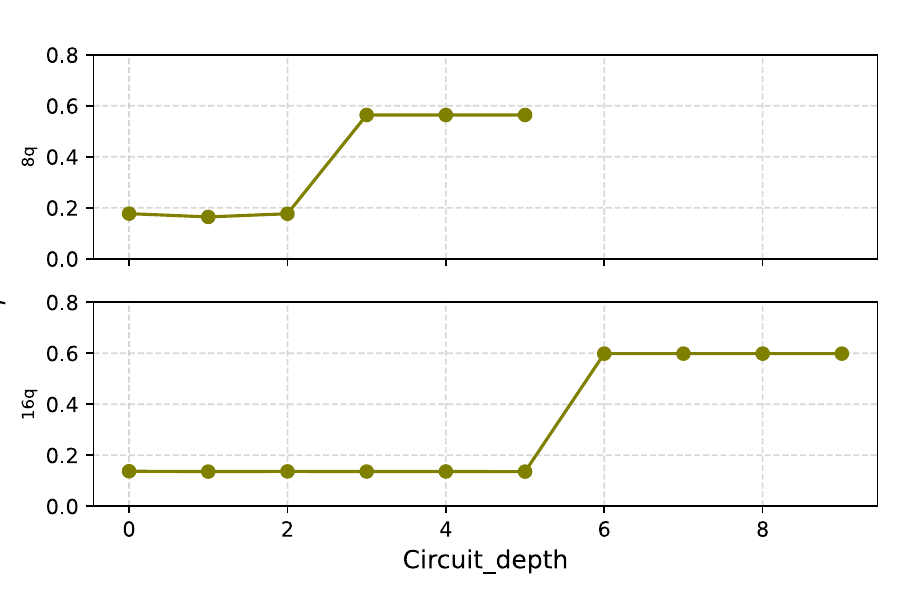}
    \caption{}
    \label{fig:causal_2}
  \end{subfigure}
  \hfill
  \begin{subfigure}[t]{0.32\textwidth}
    \includegraphics[width=\linewidth]{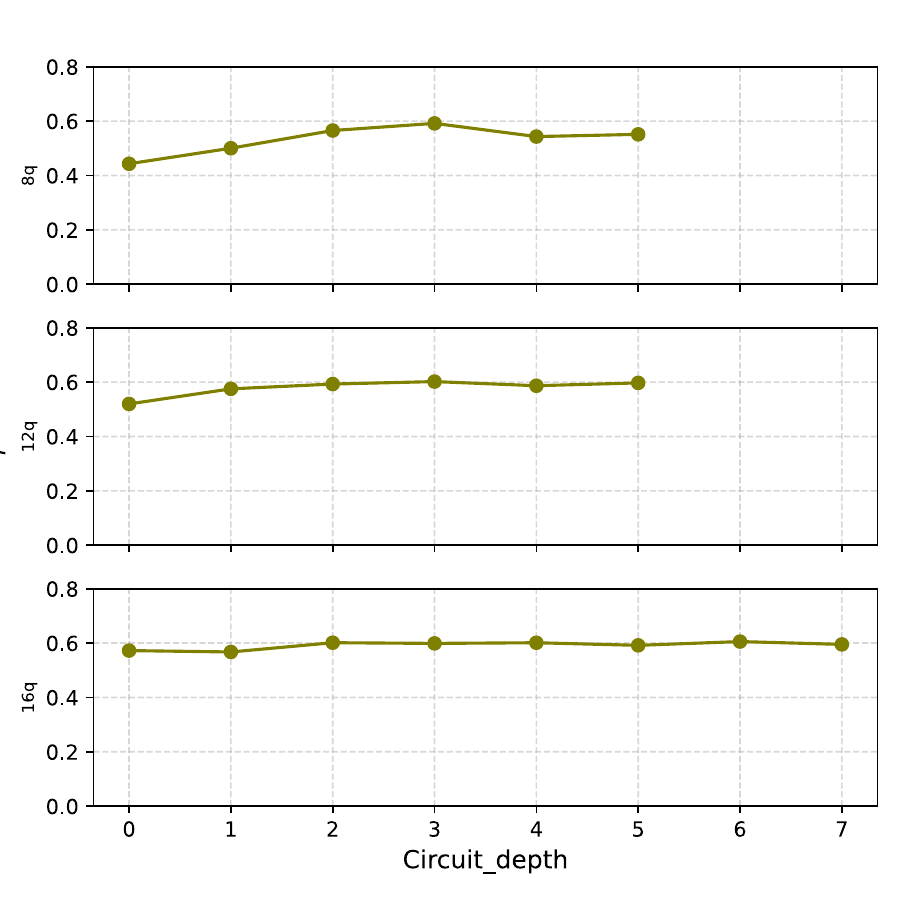}
    \caption{}
    \label{fig:causal_3}
  \end{subfigure}
  \caption{Comparison of $\tilde{r}$ values after the second $T$ gate layer for different entanglement heating architectures. Each circuit begins with $n$ $T$--states in the initialization layer and $n$ $T$ gates in the second $T$ layer. The three subfigures correspond to: (a) a causally covered random Clifford circuit, (b) a bitonic sorting network (restricted to $n=8$ and $n=16$ due to its $2^k$ structure), and (c) a Clifford circuit constructed from a random cyclic permutation. For each architecture, circuits of $n=8$, $12$, and $16$ qubits are tested (where applicable), and the results are averaged over 20 random circuit instances.For architectures (b) and (c), one unit of entanglement heating depth is defined as a pair of consecutive layers: one of CNOT gates and one of randomly chosen single--qubit Clifford gates ($H$ or $S$).}
  \label{fig:entanglement_heating}
\end{figure*}
\subsection{Causal Cover}

In this section, we introduce the notion of a \textit{causal cover}, which provides a structured yet universal framework for introducing controlled entanglement and routing in quantum circuits. This concept can be used to de--randomize chaotic circuits by imposing a deterministic combinatorial structure on the underlying interaction graph, while still achieving the high connectivity necessary for chaos and scrambling. Intuitively, a causal cover ensures that information can propagate from any qubit to any other qubit through time--respecting paths in the circuit. Formally, we define this as follows.

\textbf{Definition}:Let \( G = (V, E) \) be a graph, and let \( (G_t) \) be a sequence of graphs such that each \( G_t = (V_t, E_t) \) satisfies \( V_t = V \) and \( E_t \subseteq E \), where \( E_t \) is a matching in \( G \). For \( u, v \in V \), we say that the pair \( (u, v) \) is \emph{causally covered} if there exists a path
\[
(i_1, i_2), (i_2, i_3), \ldots, (i_{n-1}, i_n)
\]
with $i_1=u$ and $i_n=v$, such that for each edge \( (i_x, i_{x+1}) \in E_{m} \), it holds that \( (i_{x+k}, i_{x+k+1}) \in E_{n} \) with \( m < n \) for all $0 < k < n$. 

In Figure.~\ref{fig:causalExplain}, we see that $(1,4)$ are causally covered since there exists a path with edges $(1,2)\in E_1$, $(2,5) \in E_2$ and $(5,4)\in E_3$, shown by the green line in circuit. We can also see that $(4,2)$ is not causally covered as we cannot get path with edges in strictly increasing $E_t$. While (2,4) is causally covered with due to $(2,5) \in E_2$ and $(5,4)\in E_3$. So circuit as a whole is not causally covered.

\textbf{Bitonic Sorting network}: Let \( G = (V, E) \) be a complete undirected graph where \( V = \{0, 1, \ldots, n-1\} \) represents \( n \) qubits and \( E = \{(i,j) \mid i,j \in V, i \neq j\} \). A bitonic sorting network of \( n \) elements consists of a recursive sequence of \textit{compare-and-swap} operations organized into alternating \textit{compare} and \textit{merge} stages.

Let \( G_t = (V_t, E_t) \) be sequence of graph such that, where \( V_t = V \) and \( E_t \subseteq E \) represents the active compare-swap edges at depht (or layer) \( t \). Each layer in the bitonic sorting network corresponds to one such \( E_t \). Since the network has depth \( O(\log^2 n) \), we have a total of \( T = O(\log^2 n) \) graph sequence.

Because bitonic sorting networks \cite{batcher1968sorting} are designed to sort any input sequence, there exists a directed path from any vertex \( i \in V \) to any other vertex \( j \in V \) over strictly increasing time indices. Therefore, the complete graph is causally covered over graph sequence \( \{G_t\}_{t=1}^{T} \) .

\textbf{Permutation network}: Let \( G = K_n \) be the complete graph on \( n \) vertices, with vertices labeled from \( 1 \) to \( n \). For a given permutation \( \pi \in S_n \), let \( O_{\pi} \) and \( E_{\pi} \) be two perfect matchings of \( G \) that implement the two-step routing protocol \cite{alon1993routing} for \( \pi \), as introduced by Alon and Goddard.

Let \( \Pi = (\pi_1, \ldots, \pi_k) \) be a sequence of permutations, and define a stepwise path-routing protocol to realize the composed permutation \( \pi = \pi_1 \circ \pi_2 \circ \cdots \circ \pi_k \). Let, 
\[M = (O_{\pi_1},E_{\pi_1}, O_{\pi_2}, E_{\pi_2},.., O_{\pi_k},E_{\pi_k})=(M_1, M_2,.., M_{2k})\]
be the sequence of perfect matchings applied sequentially to implement the full permutation.

If we choose each \( \pi_i \in \Pi \) to be a cyclic permutation rather than a random permutation, the resulting sequence of matchings \( M \) forms a brickwork-like structure. In this arrangement, each vertex participates in local interactions with its neighbors in a regular, repeating pattern across time steps. This structure ensures the existence of time-respecting paths between all vertex pairs. As a result we get causally covered network with the use of cyclic permutation.

\section{Prior work}

Prior studies have extensively investigated the onset of quantum chaos in circuits through both entanglement spectrum statistics and out-of-time order correlators (OTOCs). The work \textit{Single T gate in a Clifford circuit drives the transition to universal entanglement spectrum statistics} ~\cite{zhou2020single} demonstrates that the addition of even a single non-Clifford gate can induce a transition toward Wigner-Dyson (WD) statistics in the entanglement spectrum, with the transition becoming more robust as the number of $T$ gates increases. \cite{haferkamp2022random} shows that the random quantum circuit will create approximate unitary t-design in depth $O(nt^{5+o(1)})$. \cite{leone2021quantum} Shows that for a quantum circuit to simulate quanatum chotic behavior, it is both necessary and suffiecient that number of t gate dopped is $\theta(n)$. \cite{leone2024learning} talks about how $k\ge2\times n$ t-gates are suffiecient to drive the circuit property to universal behaviour. 

Other works ~\cite{leone2021quantum} suggest that circuits with approximately $2n$ non-stabilizer resources are sufficient to exhibit universal entanglement properties and become learning-hard. Motivated by these results, we structure our circuits with two layers of $n$ $T$-gates. 
Meanwhile, work \textit{Information Scrambling in Computationally Complex Quantum Circuits} ~\cite{mi2021information} uses OTOC decay to quantify chaos and shows that sufficiently deep circuits with enough non-Clifford resources exhibit operator entanglement growth consistent with chaotic dynamics. These works collectively connect the emergence of chaos to the quantity of magic (non-Clifford) and the entangling architecture of the circuit. Other work \cite{garcia2022out,xu2024scrambling} worked on the Operator Spreading and OTOC in chaotic circuit. Our work builds on these insights by designing deterministic architectures with causal cover properties to minimize depth while achieving chaotic signatures in both spectral and dynamical diagnostics.

\begin{figure*}[!t]
\begin{subfigure}[t]{0.32\textwidth}
    \includegraphics[width=\linewidth]{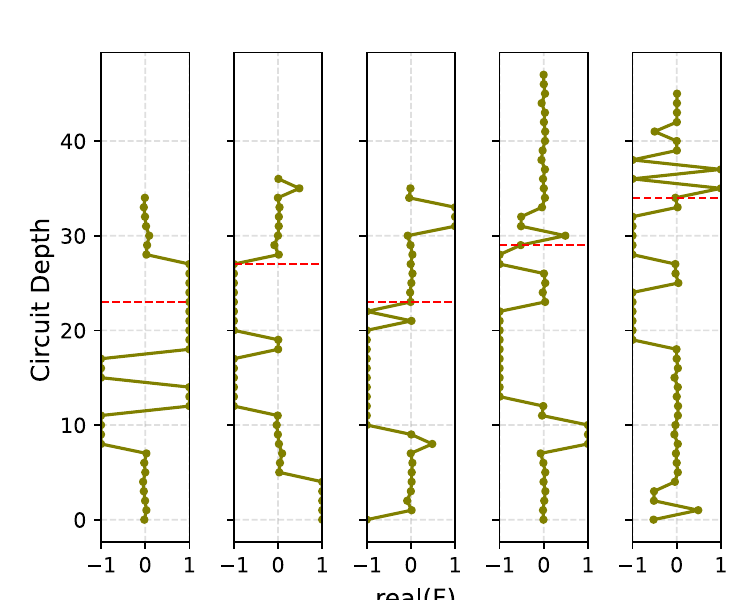}
    \caption{}
    \label{OTOC_a}
  \end{subfigure}
  \hfill
  \begin{subfigure}[t]{0.32\textwidth}
    \includegraphics[width=\linewidth]{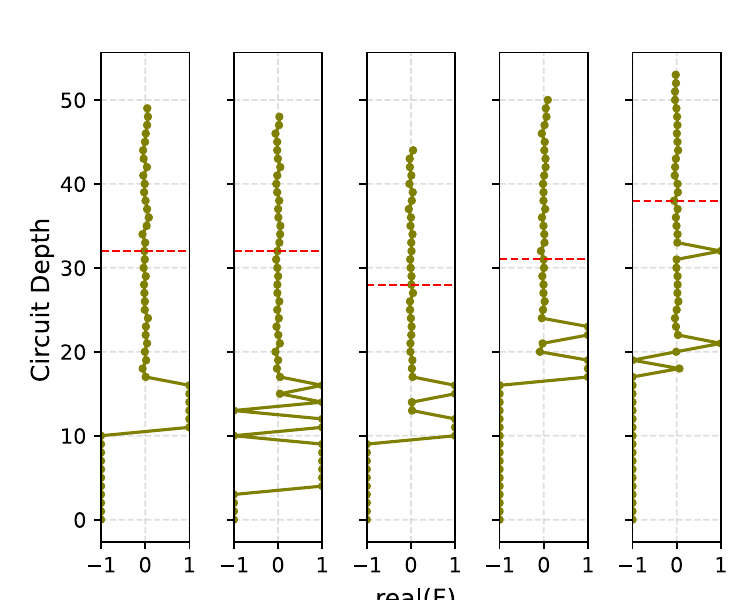}
    \caption{}
    \label{OTOC_b}
  \end{subfigure}
  \hfill
  \begin{subfigure}[t]{0.32\textwidth}
    \includegraphics[width=\linewidth]{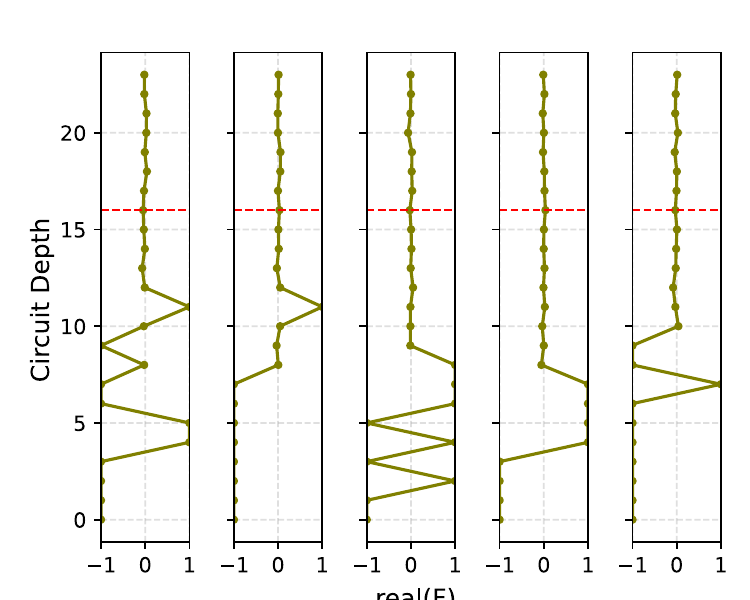}
    \caption{}
    \label{OTOC_c}
  \end{subfigure}
  \caption{OTOC decay comparison for different circuit architectures(n=20 qubits). The plot shows the OTOC value $real(F)$ as a function of circuit depth. (a) corresponds to a 4-block circuit with a random Clifford unitary as the entanglement heating section. (b) represents a 5-block architecture where the entanglement heating section is a random Clifford unitary satisfying the causal cover property. (c) also uses a 5-block structure, with the entanglement heating section constructed using a causally covered Clifford circuit based on a random permutation routing algorithm. Red dotted line denotes the depth of second layer of T-gates. The 5-block circuits show clear decay in OTOC after the second T-layer, indicating chaotic behavior, while the 4-block circuits do not consistently reach the chaotic regime.}
  \label{fig:OTOC decay}
\end{figure*}
\section{Experimental Setup}

Prior works on entanglement spectrum statistics \cite{zhou2020single} typically employ quantum circuits composed of random initial states, followed by random Clifford gates for entanglement heating, insertion of $T$ gates, and further Clifford evolution. These constructions are designed to approximate unitary $t$-designs and achieve Wigner-Dyson (WD) statistics through random circuit. In this work, we aim to \textit{derandomize}(replace random clifford circuit with structured circuit) this process by identifying and preserving the essential mechanisms that gives rise to universal statistics, while replacing stochastic elements with structured counterparts.

Our key observation is that two ingredients are critical: (i) the presence of non-stabilizerness in the initial state, and (ii) the ability of Clifford evolution to spread this non-stabilizerness throughout the system. A subsequent injection of non-Clifford resources (e.g., additional $T$ gates) then interacts with the entangled non-stabilizer content to generate complex operator structure, which is further amplified by additional Clifford evolution. This process ultimately drives the circuit toward WD spectral statistics.

Based on this, we divide our circuit architecture into four stages: \textbf{(1) Initialization}, \textbf{(2) Entanglement Heating}, \textbf{(3) Additional $T$-Layer}, and \textbf{(4) Final Entanglement Heating}. Instead of a random product state, we initialize the system using a layer of $T$ states applied to each qubit to inject structured non-stabilizerness. The intermediate $T$-layer contains a variable number of $T$ gates, depending on the experiment. The entanglement heating stages, originally the most random part of the circuit, are the primary focus of our derandomization efforts. We replace them with carefully constructed, structured unitaries as described in the next section.

\subsection{Entanglement Heating architecture}
Entanglement heating is the process of increasing the entanglement entropy of the cirucit. It has been done in previous works applying random clifford unitaries to the circuit \cite{shaffer2014irreversibility}. To remove this random factor from the circuit and to investigate the emergence of chaotic behavior and scrambling, we implement the entanglement heating section of our circuits using three distinct architectures. We construct structured yet efficient unitaries to evaluate the impact of deterministic connectivity on operator spreading and spectral statistics.

\textbf{Random Clifford Circuit with Causal Cover.} 
We iteratively apply random layers of Clifford gates drawn from \{H, S, CNOT\} to the circuit. After each layer, we verify whether the circuit satisfies the \textit{causal cover} condition—i.e., whether the entire system is interconnected such that all qubits can influence one another through the circuit's light cone. Layers are added until this condition is met.

\textbf{Random Permutation via Two-Step Routing.} 
We generate random cyclic permutations over the \( n \)-qubit register and use the two-step routing method proposed by Brierley \textit{etal.} \cite{alon1993routing} to convert each permutation into two layers of SWAP operations. These are then replaced with equivalent CNOT gate patterns to form a brickwork entangling structure. After each CNOT layer, we apply single-qubit Clifford gates (H or S) uniformly. This construction results in a brick-like circuit that is inherently causally covered due to the mixing properties of the random permutation and routing.

\textbf{Sorting Network-Based Architecture.} 
We employ the Bitonic sorting network ~\cite{batcher1968sorting}, where each compare-and-swap operation is implemented as a CNOT gate. Following each CNOT, we apply single-qubit Clifford gates (H or S) on both participating qubits. Bitonic networks guarantee full causal coverage and have circuit depth \( O(\log^2 n) \). For comparison, we also reference the AKS sorting network ~\cite{ajtai19830}, which achieves optimal \( O(\log n) \) depth but incurs an impractically large constant factor (depth $\approx 6100log(n)$), making it unsuitable for near-term experiments. The bound has been reduced over the course of time but still is impractically large depth for reasonable values of n \cite{dobrokhotova2022constant}. However, this asymptotic behavior suggests that for sufficiently large \( n \), logarithmic-depth circuits may suffice for causal connectivity and efficient entanglement spreading.
As demonstrated in Figure 1, the depth of the entanglement heating circuit does not significantly impact the emergence of Wigner-Dyson statistics, provided that the causal cover condition is satisfied.

\subsection{OTOC Experiments}

To probe information scrambling and chaotic dynamics, we employ the interferometric protocol \cite{swingle2016measuring} to measure the out-of-time-order correlator (OTOC) using real(\( F(t) \)) function for each circuit at varying depths. For purely Clifford circuits, which remain within the stabilizer formalism, the OTOC exhibits large fluctuations and typically oscillates between $-1$ and $1$, as shown in Figure.~\ref{fig:OTOC decay}. However, the inclusion of non-Clifford resources such as $T$ gates drives the system toward chaotic behavior, with \( F(t) \) decaying toward zero. A circuit is considered to exhibit quantum chaos when \( F(t) \to 0 \), indicating maximal scrambling.

For the OTOC experiment, we designed two circuit architectures. The first is a four-block structure identical to the one used in the entanglement spectrum statistics experiment. The second extends this by introducing a random Clifford layer with the causal cover property at the beginning, forming a five-block architecture. In both cases, the remaining four sections are kept the same. Our comparison reveals that the four-block architecture does not consistently exhibit chaotic behavior across all circuits. In contrast, the five-block architecture consistently reaches the chaotic OTOC regime—marked by OTOC decay—after the second T-gate layer Figure.~\ref{fig:OTOC decay}. This suggest that with 4-block architecture we are unable to get operator entanglement while with 5 block architecture we get the Operator entanglement.

Our results show that even a single causal cover layer is sufficient to drive the system into the chaotic regime when enough $T$ gates are inserted. The decay of the OTOC value corroborates the emergence of Wigner-Dyson statistics in the entanglement spectrum, confirming that our derandomized circuit designs support scrambling and chaos. Notably, with the Bitonic sorting network, we achieve chaotic behavior in depth \( O(\log^2 n) \), and with the AKS sorting network \cite{dobrokhotova2022constant}, in theoretical depth \( O(\log n) \). While AKS is impractical for small systems due to its large constant factor, these results imply that in the asymptotic limit, deterministic circuits of logarithmic depth can still replicate the hallmarks of quantum chaos.

\section{Results}

We conducted a sequence of experiments to evaluate how circuit architecture and depth affect the transition of the entanglement spectrum statistics \( \tilde{r} \) from Poisson to Wigner-Dyson (WD) distribution, a hallmark of quantum chaos and approximate unitary \( t \)-designs. In the first experiment, we tested whether increasing the depth of the entanglement heating section –while preserving the causal cover property –would improve convergence to WD statistics. Our results Figure.~\ref{fig:causal_cover} show that once the causal cover condition is satisfied, further increasing the depth yields no significant improvement in \( \tilde{r} \), suggesting that satisfying causal coverage alone is sufficient for spectral statistics to approach the WD regime. In the second experiment, we compared three distinct entanglement heating architectures: random Clifford circuits with causal cover, bitonic sorting networks, and cyclic permutation-based circuits. Despite structural differences, all architectures led to similar \( \tilde{r} \)-distributions that matched WD statistics, reinforcing the conclusion that causal cover, rather than circuit randomness or specific depth, is the critical feature. Notably, the bitonic sorting network and cyclic permutation circuit achieved these results with depths of \( O(\log^2 n) \) and \( O(\log n) \), respectively. This suggests that with careful, deterministic design, circuits of polylogarithmic depth can suffice to approximate unitary \( t \)-designs. Theoretically, use of AKS sorting network can achieve this in $O(log(n))$ depth but AKS doesn't yet have any efficient implementation. 
Our third experiment focuses on derandomizing the circuit structure to ensure consistent OTOC decay characteristic of chaotic dynamics. As shown in Figure.~(c), the four-part architecture exhibits fluctuating OTOC values for some circuit instances, indicating a lack of universal chaotic behavior. To address this, we introduce a five-part architecture by prepending a random Clifford layer with the causal cover property. This modification ensures that all circuit instances consistently display OTOC decay after the second T-layer, signaling the onset of chaos. Since the entanglement heating sections across all configurations maintain causal cover depth, the overall circuit depth scales as $O(\log n)$ theoretically.

\bibliography{citation}
\vspace{12pt}
\end{document}